\documentclass[doublecol,linenumbers]{epl2} 

\usepackage{cite}
\usepackage[utf8]{inputenc}
\usepackage{graphicx}
\usepackage{graphicx,epstopdf}
\usepackage{dcolumn}
\usepackage{amsmath}
\usepackage{lineno,hyperref}
\usepackage{amsfonts}
\usepackage{amsthm}
\usepackage{bbm}
\usepackage{amssymb}
\usepackage{mathrsfs}
\PassOptionsToPackage{caption=false}{subfig}
\usepackage{subfig}
\usepackage{color}
\usepackage[normalem]{ulem}
\usepackage{blindtext}
\usepackage{lipsum}
\usepackage{bbm}

\newcommand{\Dcal}{\mathcal{D}}

\newcommand{\Rcal}{\mathcal{R}}

\newcommand{\Fcal}{\mathcal{F}}

\newcommand{\1}{\mathbbm{1}}

\newcommand{\Rmath}{\mathbbm{R}}

\newcommand{\ket}[1]{| #1 \rangle}
\newcommand{\bra}[1]{\langle #1 |}

\newcommand{\interpro}[2]{\langle #1 | #2 \rangle}

\newcommand{\ketus}[0]{|\!\! \uparrow \rangle}
\newcommand{\ketds}[0]{|\!\! \downarrow \rangle}

\newcommand{\trs}[1]{ \text{Tr} \{ #1 \}}

\title{Optimizing NMR quantum information processing via generalized transitionless quantum driving}
\shorttitle{Optimizing NMR quantum information processing via generalized transitionless quantum driving} 

\author{Alan C. Santos\inst{1,*}, Amanda Nicotina\inst{2}, Alexandre M. Souza\inst{2}, Roberto S. Sarthour\inst{2}, \\Ivan S. Oliveira\inst{2}, Marcelo S. Sarandy\inst{1} }
\shortauthor{Alan C. Santos}

\institute{                    
	\inst{1} Instituto de F\'{i}sica, Universidade Federal Fluminense, Av. Gal. Milton Tavares de Souza s/n, Gragoat\'{a},\\ 24210-346 Niter\'{o}i, Rio de Janeiro, Brazil \\
	\inst{2} Centro Brasileiro de Pesquisas Físicas, Rua Dr. Xavier Sigaud 150, 22290-180 Rio de Janeiro, Brazil
	\\ \inst{*}Electronic address: ac\_santos@id.uff.br
}


\abstract{High performance quantum information processing requires efficient control of undesired decohering effects, which are present in realistic quantum dynamics. 
	To deal with this issue, a powerful strategy is to employ transitionless quantum driving (TQD), where additional fields are added to speed up the 
	evolution of the quantum system, achieving a desired state in a short time in comparison with the natural decoherence time scales. In this paper, 
	we provide an experimental investigation of the performance of a generalized approach for TQD to implement shortcuts to adiabaticity in nuclear 
	magnetic resonance (NMR). As a first discussion, we consider a single nuclear spin-$\frac{1}{2}$ system in a time-dependent rotating magnetic field. 
	While the adiabatic dynamics is violated at a resonance situation, the TQD Hamiltonian is shown to be robust against resonance, 
	allowing us to mimic the adiabatic behavior in a fast evolution even under the resonant configurations of the original (adiabatic) Hamiltonian. Moreover, we show that the generalized TQD theory 
	requires less energy resources, with the strength of the magnetic field less than that required by standard TQD. As a second discussion, we analyze the experimental 
	implementation of shortcuts to single-qubit adiabatic gates. By adopting generalized TQD, we can provide feasible time-independent driving Hamiltonians, 
	which are optimized in terms of the number of pulses used to implement the quantum dynamics. 
	The robustness of adiabatic and generalized TQD evolutions against typical decoherence processes in NMR is also analyzed.}

\begin{document}

\maketitle

\section{Introduction} Adiabatic dynamics play an important role in a number of applications in quantum mechanics (see, e.g.,~Ref.~\cite{Tameem:18}). Under the effect of a surrounding environment, 
the success of an adiabatic protocol depends on the time scale required by the quantum evolution in comparison with the relaxation time scale 
of the external environment~\cite{Sarandy:05-1,Venuti:17,Albash:15}. In this context, a powerful strategy for minimizing the reservoir effects is provided by 
shortcuts to adiabaticity. \textit{Transitionless quantum driving} (TQD)~\cite{Demirplak:03,Demirplak:05} 
has been a protocol widely used for speeding up the adiabatic dynamics. Differently from other approaches~\cite{Santos:18-a,Li:18,Huang:17}, 
TQD evolutions allow us to exactly mimic the adiabatic dynamics through additional fields, which inhibit transitions between the energy levels of the original 
(adiabatic) Hamiltonian. TQD has been used in different scenarios in quantum physics, such as quantum thermodynamics~\cite{Funo:17,Adolfo:14,Abah:17}, 
quantum computation and information~\cite{Santos:15,Santos:16}, and state engineering~\cite{Santos:18-b,Hu:18}.

The main idea behind TQD is to add an auxiliary term $H_{\text{cd}}(t)$, which is called counter-diabatic Hamiltonian, to the original Hamiltonian $H_{0}(t)$, 
so that we get the transitionless driving Hamiltonian $H_{\text{tqd}}(t)$. In general, the design of TQD requires  
highly non-local terms in the counter-diabatic Hamiltonian $H_{\text{cd}}(t)$~\cite{Adolfo:13-Shortcut,Muga:10}, as well as strong time dependence on 
the additional fields~\cite{Santos:16}. These issues typically make transitionless Hamiltonians hard to be implemented in realistic physical 
systems~\cite{Xia:16}. Here, we are interested in investigating the performance of a  
generalized approach of TQD, which can be taken as a wider approach for the optimization of TQD protocols. More specifically, we analyze the 
implementation of generalized TQD in a nuclear magnetic resonance (NMR) experimental setup. In particular, we realize feasible time-independent transitionless 
Hamiltonians~\cite{Santos:18-b}, which provide the optimal dynamics concerning both their experimental viability 
and energy cost, as measured by the strength of the counter-diabatic fields required by the implementation~\cite{Hu:18}. 
As a first experiment, we consider a single nuclear spin-$\frac{1}{2}$ system in a time-dependent rotating magnetic field, analyzing the effects of the shortcut 
to the adiabaticity at a resonance situation. As a second discussion, we analyze the experimental implementation of shortcuts to single-qubit adiabatic gates. 
For each evolution we derive the standard TQD Hamiltonian, as provided by standard TQD~\cite{Demirplak:03,Demirplak:05}, and the optimal 
TQD Hamiltonian obtained from the generalized TQD protocol~\cite{Santos:18-b}. The robustness of both the adiabatic and generalized TQD evolutions against typical 
decoherence processes in NMR are also analyzed.

\section{Shortcuts to adiabaticity through generalized TQD }

Let us consider a discrete quantum system {\it S} defined in a $\Dcal$-dimensional Hilbert space ${\cal H}$. A generalized TQD for {\it S} is obtained from the evolution operator~\cite{Santos:18-b}
\begin{align}
U(t) = \sum_{n = 1}^{\Dcal}\nolimits e^{i \int_{0}^{t}\theta_{n}(\xi)d\xi} \ket{n(t)}\bra{n(0)} ,
\end{align}
where $\{\ket{n(t)}\}$ is the set of eigenstates of a reference Hamiltonian $H_{0}(t)$. In general, $H_{0}(t)$ is identified as a slowly piecewise time-dependent 
Hamiltonian, so that the adiabatic dynamics can be achieved. The parameters $\theta_{n}(t)$ are real arbitrary functions to be adjusted in order to optimize some 
physical relevant quantity~\cite{Santos:18-b} or, as we shall see, optimize some pulse sequence for achieving some output state in NMR quantum information 
processing. From a general definition of $U(t)$, the driving generalized transitionless Hamiltonian $H^{\text{gen}}_{\text{tqd}}(t)=-i \hbar U(t) \dot{U}^{\dagger}(t)$  reads~\cite{Santos:18-b}
\begin{align}
H^{\text{gen}}_{\text{tqd}}(t) &= i \hbar \sum_{n}\nolimits\ket{\dot{n}\left(t\right)}\bra{n\left( t\right)}+i\theta _{n}\left( t\right)\ket{n\left(t\right)}\bra{n\left( t\right)} . \label{GenHSA}
\end{align}
In particular, we can get the \textit{standard} TQD protocol as provided by Demirplak and Rice~\cite{Demirplak:03,Demirplak:05} and by Berry~\cite{Berry:09}, 
if we particularize $\theta_{n}(t) = \theta^{\text{ad}}_{n}(t)$, where
$\theta^{\text{ad}}_{n}(t) = - E_{n}(t)/\hbar - i\interpro{\dot{E}_{n}(t)}{E_{n}(t)}$
is the phase that accompanies the adiabatic dynamics. In this case, it is possible to show that the standard transitionless Hamiltonian, which drives the system, is given by
\begin{eqnarray}
H^{\text{std}}_{\text{tqd}}(t) = H_0(t) + H_{\text{cd}}(t) \text{ , }
\label{HTtqd}
\end{eqnarray}
where $H_{\text{cd}}(t)$ is the \textit{counter-diabatic} term added to $H_0(t)$ to supress the diabatic transitions that would typically arise during the evolution. 
The Hamiltonian $H_{\text{cd}}(t)$ reads
\begin{eqnarray}
H_{\text{cd}}(t) = i \hbar \sum_{n}\nolimits\ket{\dot{n}\left(
	t\right)}\bra{n\left( t\right)}+ \gamma_{n}(t)\ket{n(t)}\bra{n\left( t\right)} ,
\end{eqnarray}
where we have defined $\gamma_{n}(t)=\interpro{\dot{n}(t)}{n(t)}$. The set $\{\theta _{n}(t)\}$ of free parameters in $H^{\text{gen}}_{\text{tqd}}(t)$ allows us to design different Hamiltonians to the same shortcut proposal. 
In particular, these parameters have been used to minimize the energy cost $\Sigma (\tau)$ of a TQD as measured by the Hamiltonian norm, which has been experimentally explored in Ref.~\cite{Santos:18-b}.
The energy cost has also been directly considered in terms of the intensity of the fields used to implement the Hamiltonian, as experimentally explored in Ref.~\cite{Hu:18}.  
Here, we are interested in theoretically and experimentally investigating how to control couplings and fields to enhance and realize implementation costs as provided by 
the NMR field intensity and pulse sequence. In particular, as a first contribution of this work, we present an experimental realization in NMR that explicitly shows that 
the energy cost as provided by field intensities can be minimized by suitable fast evolutions.
Moreover, as a second contribution exhibiting the usefulness of $\{\theta _{n}\left( t\right)\}$, we further study the applicability of generalized TQD to accelerate the implementation 
of adiabatic gates. To this end, we study the pulse sequence required to different implementations: adiabatic dynamics, standard TQD, and generalized TQD. As we shall see, while the 
adiabatic dynamics and the standard TQD require a long pulse sequence (due to Trotterization), generalized TQD allows us to provide an enhanced pulse sequence, 
which does not requires Trotterization. In this sense, the free parameters $\theta _{n}\left( t\right)$ can be used to minimize the digital pulse sequence necessary to 
implement quantum gates.


\section{Optimal TQD for a single-spin in a rotating magnetic field}
As a first application of generalized TQD, we consider a single spin-$\frac{1}{2}$ particle in the presence of a rotating magnetic field 
$\vec{B}_{0}(t)\!=\!B_{z} \hat{z} + B_{xy} (\cos \omega t \hat{x}+\sin \omega t \hat{y})$, with $B_z, B_{xy}, \omega \in \Rmath$. The Hamiltonian $H_{0}(t)\!=\!- \gamma \vec{S} \cdot \vec{B}_{0}(t)$ associated 
with the system then reads~\cite{Sarthour:Book}
\begin{eqnarray}
H_{0}(t) = (\hbar/2) \left[ \omega_{z}\sigma_{z} + \omega_{xy} (\sigma_{x}\cos \omega t + \sigma_{y}\sin \omega t ) \right] , \label{H0}
\end{eqnarray} 
where the frequencies $\omega_{z}$ and $\omega_{xy}$ are given by $\omega_{z} = - \gamma B_{z}$ and $\omega_{xy} = - \gamma B_{xy}$, respectively, 
with $\gamma$ denoting the gyromagnetic ratio and $\omega$ the rotating frequency of a 
radio-frequency (RF) field. In particular, it is well-known that the adiabatic behavior of the Hamiltonian $H_{0}(t)$ is drastically affected 
in resonant regimes ($\omega \approx \omega_{z}$)~\cite{Suter:08}, where undesired transitions between ground $\ket{E_{0}(t)}$ and 
excited $\ket{E_{1}(t)}$ states are induced due to resonance. From Eq.~\eqref{GenHSA}, we can write the Hamiltonian of the generalized 
transitionless evolution for $H_{0}(t)$ as $H^{\text{gen}}_{\text{tqd}}(t)\!=\!- \gamma \vec{S} \cdot \vec{B}^{\text{gen}}_{\text{tqd}}(t)$, with the magnetic field given by ($\alpha = \arctan( \omega_{xy}/\omega_{z} )$)
\begin{eqnarray}
\vec{B}^{\text{gen}}_{\text{tqd}}(t) = B_{z}^{\theta} \hat{z} + B_{\theta}(t)\sin \alpha (\cos \omega t \hat{x}+\sin \omega t \hat{y}) , \label{GenBSA}
\end{eqnarray}
where we define $\theta(t) = \theta_{0}(t) - \theta_{1}(t)$, $B_{z}^{\theta}\!=\!B_{\omega} + B_{\theta}(t)\cos \alpha$, $B_{\theta}(t)\!=\!- \theta(t)/\gamma$, $B_{\omega}\!=\!- \omega/\gamma$. Note that a transitionless quantum 
driving is not obtained by changing the angular frequency of the RF field, but rather by changing the intensity of the magnetic fields. Therefore, the generalized transitionless theory 
allows us to use the phases $\theta_{n}(t)$ ($n\!=\!0,1$) to minimize the magnetic field required to implement the desired dynamics. In some cases, such a field 
is fixed and no free parameter can be used to optimize the field intensity. For example, if we set the parameters $\theta_{n}(t)$ in order to obtain 
the exact adiabatic dynamics with adiabatic phases $\theta^{\text{ad}}_{n}(t)$, we find the standard shortcut to adiabaticity, which is provided by the 
Hamiltonian in Eq.~(\ref{HTtqd}) and reads as $H^{\text{std}}_{\text{tqd}}(t)\!=\!-\gamma \vec{S} \cdot \vec{B}^{\text{std}}_{\text{tqd}}(t)$, with the magnetic field $\vec{B}^{\text{std}}_{\text{tqd}}(t)$ to implement $H^{\text{std}}_{\text{tqd}}(t)$ reading
\begin{eqnarray}
\vec{B}^{\text{std}}_{\text{tqd}}(t) = B^{\text{std}}_{z}\hat{z} + B^{\text{std}}_{xy}(\cos \omega t \hat{x}+\sin \omega t \hat{y}),
\label{eqBTtqd}
\end{eqnarray}
where $B^{\text{std}}_{xy}\!=\!B_{xy} - \sin (2\alpha) B_{\omega}/2 $ and $B^{\text{std}}_{z}\!=\!B_{z} + B_{\omega}\sin^2\alpha$, with $B_{\omega}$ denoting the additional magnetic field.
By analyzing the energy resources to implement the desired evolution, we can find the optimal protocol for the transitionless dynamics in terms of 
the magnetic field intensities required by the Hamiltonian. In particular, it is possible to show that the optimal field is obtained by setting 
$\theta_{n}(t) = i\interpro{\dot{E}_{n}(t)}{E_{n}(t)}$~\cite{Santos:18-b,Hu:18}. Therefore, for the Hamiltonian in Eq.~\eqref{H0}, the optimal transitionless counterpart reads $H^{\text{opt}}_{\text{tqd}}(t)\!=\!- \gamma \vec{S} \cdot \vec{B}^{\text{opt}}_{\text{tqd}}(t)$, with the optimal magnetic
\begin{align}
\vec{B}^{\text{opt}}_{\text{tqd}}(t) = B_{\omega}\sin^2\alpha \hat{z} - B^{\text{opt}}_{xy} (\cos \omega t \hat{x}+\sin \omega t \hat{y}) .
\end{align}
where $B^{\text{opt}}_{xy}\!=\! (B_{\omega}/2)\sin 2\alpha $. We can see that the norms of the fields satisfy $||\vec{B}^{\text{std}}_{\text{tqd}}(t)|| > ||\vec{B}_{0}(t)||$ and 
$||\vec{B}^{\text{std}}_{\text{tqd}}(t)|| > ||\vec{B}^{\text{opt}}_{\text{tqd}}(t)||$ for any choice of the set of parameters 
$\Omega = \{\omega,\omega_{z},\omega_{xy}\}$. This means that a fast evolution based on the standard shortcut 
exactly mimicking the adiabatic phase, such as given by Eq.~(\ref{HTtqd}), always requires more energy resources than the 
original adiabatic dynamics. On the other hand, the relation between $||\vec{B}^{\text{opt}}_{\text{tqd}}(t)||$ and $||\vec{B}_{0}(t)||$ 
depends on the parameters $\omega$, $\omega_{z}$ and $\omega_{xy}$ used in $||\vec{B}_{0}(t)||$. In particular, 
this means that the optimal shortcut to adiabaticity can be faster while spending even less resources 
than the original adiabatic dynamics. More specifically, this trade-off can be expressed through the norm relation
\begin{align}
\frac{||\vec{B}_{0}(t)||}{||\vec{B}^{\text{opt}}_{\text{tqd}}(t)||} = \frac{B_{xy}^2 + B_{z}^2}{B_{xy}B_{\omega}}  =  \frac{\omega_{xy}^2 + \omega_{z}^2}{\omega_{xy} \omega} 
. \label{FieldsRel}
\end{align}
For the case $\omega_{xy}\!=\!\omega_{z}\!=\!\omega$, we get $||\vec{B}_{0}(t)||\!=\!2 ||\vec{B}^{\text{opt}}_{\text{tqd}}(t)||$. This shows 
the advantage of the optimal TQD approach in comparison with its reference counterpart, whose implementation requires a higher magnetic field. It is worth mentioning that, since the experimental implementation of the driven Hamiltonian in NMR is obtained in the rotating frame, the dynamics here is implemented by a phase modulated pulse with constant amplitude applied off resonance. This approach is similar to the adiabatic pulse technique where large swept-frequency offsets are used to make the effective magnetic field, observed in the rotating frame,  approximately collinear with the spin magnetization~\cite{Garwood:01}. In our experiment no swept-frequency has been used. The energy cost is well defined from the transverse RF field used to drive the system, since the $Z$ component of the Hamiltonian is implemented by a frequency shift, with no real cost associated with it. Indeed, as shown in Supplemental Material, an advantage for the optimal TQD approach with respect to its adiabatic version is also achieved in terms of the cost of the RF field.

\begin{figure}[t!]
	\centering
	\includegraphics[scale=0.58]{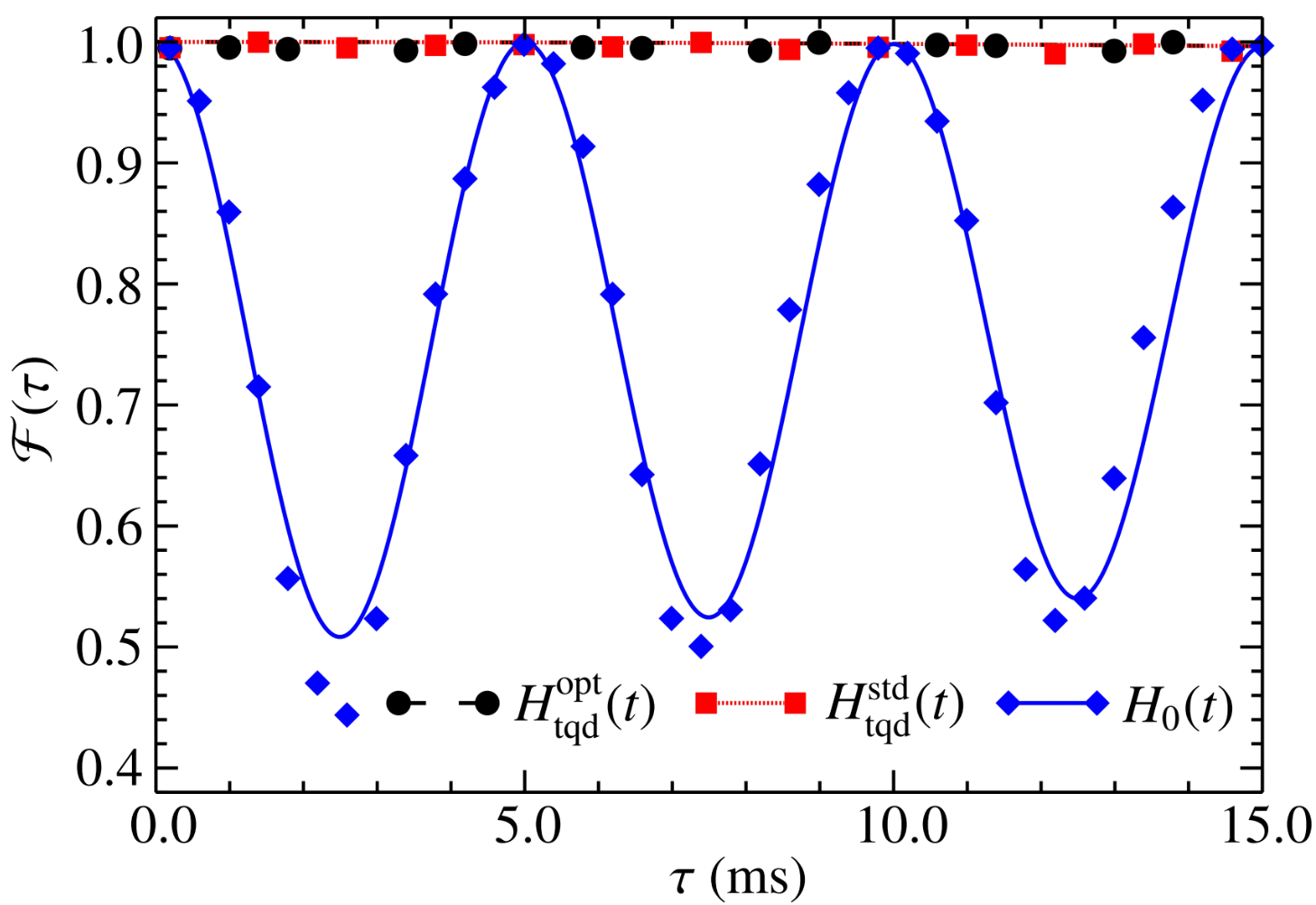}
	\vspace{-0.3cm}
	\caption{Fidelity between the instantaneous ground state of $H_0(t)$ and the dynamically evolved states driven by the Hamiltonians $H_0(t)$, $H^{\text{opt}}_{\text{tqd}}(t)$, and $H^{\text{std}}_{\text{tqd}}(t)$. Curves denote the theoretical predictions and the symbols are the experimental values. The implementation of $H_{0}(t)$ is more sensitive to RF miscalibration, so that we have less tolerance to field inhomogeneity compared to the other cases. Then, we may have a fluctuation between experimental and theoretical results in some points.}
	\label{NMR-Fidelity}
\end{figure}

In order to demonstrate the advantage of  the TQD approach as a tool in quantum control, we have performed an experimental realization 
using a two-qubit NMR system, namely, the $^1$H and $^{13}$C  spin$-1/2$ nuclei in the Chloroform molecule, with up and down spin states denoted by 
$\ketus$ and $\ketds$, respectively~\cite{Sarthour:Book}. 
The experiment has been realized at room temperature in a Varian 500 MHz spectrometer. More specifically, we have compared the 
dynamics of the system driven by fields $\vec{B}_{0}(t)$,~$\vec{B}^{\text{std}}_{\text{tqd}}(t)$ and~$\vec{B}^{\text{opt}}_{\text{tqd}}(t)$. Since we need a single qubit, the $^{1}$H spin has been driven by phase modulated magnetic fields,
while the $^{13}$C spin has been decoupled during the experiment (for more details, see supplemental material). In the rotating frame, the Hamiltonian associated with the magnetic pulse reads
\begin{eqnarray}
H_{\text{rf}} = (\hbar/2) \left[ \omega_{z}\sigma_{z} + \omega_{xy} (\sigma_{x}\cos \phi + \sigma_{y}\sin \phi ) \right] , \label{H02}
\end{eqnarray}
with $\phi$ being the angle in xy plane between $x$-axis and the direction of the radio-frequency pulse. This angle can be freely chosen and continuously varied during the experiment. By changing the amplitude of the pulses, the offset frequency and using time phase modulation, one can reproduce the desired dynamics. The evolution has been implemented in an on-resonance condition  
$\omega_z\!=\!\omega_{xy}\!=\!\omega\!=\!2\pi \times 200$Hz. This means that, even though the evolution obeys the adiabatic condition, 
it can lead to population transfer, resulting in a violation of the expected adiabatic evolution as a resonance effect~\cite{Suter:08}. 
From the values set for $\omega$, $\omega_{z}$ and $\omega_{xy}$, Eq.~\eqref{FieldsRel} implies that $||\vec{B}_{0}(t)|| = 2||\vec{B}^{\text{opt}}_{\text{tqd}}(t)||$, 
so that $||\vec{B}^{\text{opt}}_{\text{tqd}}(t)||\!<\!||\vec{B}_{0}(t)||\!<\!||\vec{B}^{\text{std}}_{\text{tqd}}(t)||$ (See Supplemental Material). 
Notice that the optimal TQD indeed spends less energy resources as measured by the strength of the magnetic field applied. 
The system is then initially prepared in the ground state of $H_{0}(0)$, given by $\ket{\psi(0)} = \cos\alpha \ketds -\sin\alpha \ketus$, and evolved according to the desired Hamiltonian. 
The quantum state experimentally obtained is determined via quantum state tomography~\cite{Leskowitz:04} 
and compared to the theoretically evaluated instantaneous ground state of $H_{0}(t)$. 
This is performed by computing the fidelity $\Fcal(t)$ between the instantaneous ground state of $H_0(t)$, represented by the density operator $\rho_{\text{gs}}(t)$, and 
the dynamically evolved quantum state driven by the Hamiltonians $H_0(t)$, $H^{\text{opt}}_{\text{tqd}}(t)$, and $H^{\text{std}}_{\text{tqd}}(t)$,
represented by the density operator $\rho(t)$. The fidelity $\Fcal(t)$ is defined here as the \textit{relative purity}~\cite{Adolfo:13-QSL}
\begin{eqnarray}
\Fcal(t) = |\trs{\rho_{\text{gs}}(t)\rho(t)}| / \left[\trs{\rho_{\text{gs}}^2(t)}\trs{\rho^2(t)}\right]^{1/2} . \label{fidelity1}
\end{eqnarray} 
The theoretical and experimental results are shown in Fig.~\ref{NMR-Fidelity} for the three distinct dynamics. 
Since the experiment is performed on-resonance, the evolution driven by the Hamiltonian  $H_{0}(t)$ is found to be 
nonadiabatic, oscillating as a function of time. On the other hand, the shortcuts given by $H^{\text{opt}}_{\text{tqd}}(t)$, 
and $H^{\text{std}}_{\text{tqd}}(t)$ keep a non-transitional evolution, being immune to the resonance effect. 

The fitting of the experimental data has been performed in an open quantum system scenario, which is accomplished by adjusting the parameters of the Lindblad equation $\dot{\rho}\!=\!\Rcal[\rho]$, with $\Rcal[\bullet]$ given by~\cite{Lindblad:76,Petruccione:Book,Alicki:Book1}
\begin{equation}
\Rcal[\bullet] =  \left[H(t), \bullet\right]/\hbar + \sum_{k}\nolimits L_{k}\bullet L_{k}^{\dagger} - \{L_{k}^{\dagger}L_{k},\bullet \}/2  ,
\label{lindbald}
\end{equation}
where $H(t)$ denotes the effective Hamiltonian of the system, which drives the unitary contribution of the dynamics, and $\{L_{k}\}$ is the set of Lindblad operators 
governing the non-unitary part of the evolution. The NMR system considered in this work is mainly affected by dephasing, 
which occurs due to the inhomogeneity of the static magnetic field. This variation causes the spins of all the molecules to slowly desynchronize and, 
therefore, lose coherence across the sample. In this case, we have a single Lindblad operator $L_{1}(t) = \gamma_{0}\sigma_{z}$, where the decohering rate 
$\gamma_{0}$ is given by $\gamma_{0} = \sqrt{1/T_2}$, with $T_2$ denoting the dephasing relaxation time of the system. 
Other non-unitary effects, such as generalized amplitude damping characterized by a relaxation scale $T_1$, are also present, but they are negligible for the time 
ranges considered in the experiment. Indeed, in our realization, we have $T_2 = 0.25$~s and $T_1 = 5.11$~s. Notice then that $T_1$ is 
two orders of magnitude 
larger than the total time of evolution, so that generalized amplitude damping does not exhibit significant effect. As previously mentioned, due to the resonance configuration, the adiabatic dynamics for $H_{0}(t) $ does not hold in general, being 
achieved just for some particular instants of time. Notice that the oscillation in the adiabatic fidelity is damped as a function of time. This is a by-product of dephasing 
in the system. Moreover, there are also errors induced by the implementation of RF pulses. Those errors are about $3\%$ per pulse. 
Remarkably, shortcuts to adiabaticity show considerable robustness against the decoherence and unitary errors. As shown in Fig.~\ref{NMR-Fidelity}, fidelity stays close to one 
throughout the evolution, which means that the states of the evolution are, indeed, kept as instantaneous eigenstates of $H_{0}(t)$. 


\section{Adiabatic single-qubit gates in NMR via Generalized TQD}
Adiabatic controlled evolutions allow for the implementation of an arbitrary quantum gate $U$ over an input state $|\psi\rangle$~\cite{Hen:15}. 
This can be accomplished by adding an auxiliary quantum system, whose dynamics induces a general  $n$-qubit gate over 
$|\psi\rangle$~\cite{Santos:15}. Here, we will illustrate the implementation of an adiabatic single-qubit gate and its acceleration 
through a generalized shortcut to adiabaticity in an NMR setup. 

To implement a single-qubit unitary transformation, we begin by considering a Hilbert space 
$\mathcal{S} = \mathcal{T} \otimes \mathcal{A} $ composed by a target subspace $\mathcal{T}$, where the gate is designed to be applied, and 
an auxiliary subspace $\mathcal{A}$. Each subspace is assumed to be associated with a single qubit. The auxiliary qubit is a fundamental element in this approach. 
In fact, in order to provide a universal scheme for adiabatic quantum computing, we need to be able to apply an arbitrary gate $U$ on any arbitrary input state. 
In addition, from the standard approach for adiabatic quantum computation, such an input state should be encoded into the ground state of an initial Hamiltonian $H$. 
Thus, given a single-qubit arbitrary state $\ket{\psi}$, the single-qubit Hamiltonian that admits $\ket{\psi}$ as ground state is $H_{\psi} = -\omega \hbar \ket{\psi}\bra{\psi}$. Therefore, the Hamiltonian is $\psi$-dependent and, by assuming $\ket{\psi}$ as an unknown state, $H_{\psi}$ cannot by determined. For this reason, we consider an auxiliary qubit 
$\mathcal{A}$ and an initial Hamiltonian $H(0)\!=\!\1_{\mathcal{T}} \otimes H_{\mathcal{A}}$, so that $\ket{E_{0}}_\mathcal{A}$ is the ground state of $H_{\mathcal{A}}$. By adopting this choice, the whole system initial state is $\ket{\Psi(0)}\!=\!\ket{\psi}_{\mathcal{T}} \otimes \ket{E_{0}}_\mathcal{A}$ for any input state $\ket{\psi}$. Here we assume $H_{\mathcal{A}}\!=\!-\hbar \omega \sigma_{z}$ and $\ket{E_{0}}\!=\!\ket{0}$.

Notice that a generic 
single-qubit quantum gate $U$ can be interpreted as a rotation of $\phi$ from the initial state $|\psi\rangle$ to the target state $|\psi_{\textrm{rot}}\rangle$ 
around an arbitrary direction $\hat{r}$ defining a state $|n_{+}\rangle$ in the Bloch sphere as
\begin{align}
|n_{+}\rangle = \cos \left(\varepsilon/2\right)\ket{0} + e^{i\delta}\sin \left(\varepsilon/2\right)\ket{1} .
\end{align}

Then, the Hamiltonian used to implement the desired operation $U$ reads
\begin{align}
H\left( s\right) = P_{\hat{r}_{+}}\otimes H^{0}\left( s\right) +P_{\hat{r}_{-}}\otimes
H^{\phi }\left( s\right) , \label{AdGate}
\end{align}%
where $H^{0}( s) $ and $H^{\phi }( s) $ are obtained from
\begin{align}
H^{\xi }( s)\! =\!-\hbar \omega [ \cos (\pi s)
\sigma _{z}+\sin (\pi s) \sigma_{xy}(\xi) ] , \label{Hxi}
\end{align}%
with $\sigma_{xy}(\xi)\! =\!\sigma _{x}\cos \xi +\sigma
_{y}\sin \xi $, $\xi\!=\!0$ and $\xi\!=\!\phi $, respectively, and $s\!=\!t/\tau$ is the dimensionless normalized time, with $\tau$ being the total evolution time. 
The information about the gate to be implemented is encoded in the parameter $\phi$ and in the projectors 
$P_{\hat{n}_{\pm}}\!=\!\ket{\hat{n}_{\pm}}\bra{\hat{n}_{\pm}}\!=\!1/2( \1 \pm \hat{r} \cdot \vec{\sigma})$, which act on $\mathcal{T}$. 
Thus, by assuming an adiabatic closed-system evolution, the composite system evolves from $\ket{\Psi(0)}\!=\!\ket{\psi}\otimes \ket{0}$ to 
$\ket{\Psi(1)}=(U\ket{\psi})\otimes \ket{1}$. 

Now, by considering the Hamiltonian for a generalized transitionless evolution as in Eq.~(\ref{GenHSA}), we can show that the gate $U$ 
can be implemented through
\begin{align}
H_{\text{tqd}}^{\text{gen}}\left( s\right) = P_{\hat{r}_{+}}\otimes H_{\text{tqd}}^{\text{gen},0}\left( s\right) +P_{\hat{r}_{-}}\otimes
H_{\text{tqd}}^{\text{gen},\phi }\left( s\right) , \label{GenGate}
\end{align}
with (See Supplemental Material)
\begin{align}
H_{\text{tqd}}^{\text{gen},\xi }(s) = 
\hbar \Xi^{\xi}_{x} \sigma_{x} - 
\hbar \Xi^{\xi}_{y}\sigma_{y} -  \hbar \Theta^{\xi} \cos (\pi s) \sigma_{z} ,
\end{align}
where we defined $\Xi^{\xi}_{x}(s)\!=\!\Theta^{\xi}(s) \sin (\pi s )\cos \xi - (\pi/\tau) \sin \xi$, $\Xi^{\xi}_{y}(s)\!=\!\Theta^{\xi}(s) \sin (\pi s) \sin \xi - (\pi/\tau) \cos \xi$ and $\Theta^{\xi}(s)\!=\!\theta^{\xi}_{0} - \theta^{\xi}_{1}$, with $\theta^{\xi}_{n}$ denoting the quantum phase accompanying the $n$-th 
eigenstate of the Hamiltonian $H^{\xi}\left( s\right)$. In particular, we can find the standard transitionless Hamiltonian by choosing 
$\theta^{\xi}_{n}(s) = \theta^{\text{ad},\xi}_{n}(s)$. In this case, we get
\begin{align}
H_{\text{tqd}}^{\text{std}}\left( s\right) = P_{\hat{r}_{+}}\otimes H_{\text{T}}^{0} +
P_{\hat{r}_{-}}\otimes H_{\text{T}}^{\phi}(s) , \label{TradGate}
\end{align}
with $H_{\text{T}}^{\xi} = H^{\xi}(s) + H_{\text{cd}}^{\xi}$ and where each counter-diabatic term is time-independent and given by 
\begin{align}
H_{\text{cd}}^{\xi} = - (\hbar \pi/\tau) \left( \sin \xi \sigma_{x} - \cos \xi \sigma_{y}\right).
\end{align} 
Notice then that the non-adiabatic transitions in adiabatic quantum computation as provided by the Hamiltonian in Eq.~\eqref{AdGate} can be 
suppressed by time-independent quantum control.
More generally, the Hamiltonian in Eq.~\eqref{GenGate} provides a remarkabe flexibility, since we can mimic the adiabatic dynamics using an infinite class of Hamiltonians. 
In particular, the gauge freedom for the parameters $\theta^{\xi}_{n}(t)$ in Eq.~\eqref{GenGate} allows, e.g. for a time-independent total Hamiltonian 
$H_{\text{tqd}}^{\text{gen},\xi }\left( s\right)$. Indeed, the optimal choice with respect to both energy cost and 
experimental feasibility can be obtained by setting $\theta^{\xi}_{n}(s) = -(i/\tau)\interpro{d_{s}E_{n}^{\xi}(s)}{E_{n}^{\xi}(s)}$~\cite{Santos:18-b}, where $\ket{E_{n}^{\xi}(s)}$ 
denotes the eigenvectors of the Hamiltonian in Eq.~\eqref{Hxi}. Then, the optimal Hamiltonian reads
\begin{align}
H_{\text{tqd}}^{\text{opt}} = P_{\hat{r}_{+}}\otimes H_{\text{cd}}^{0} +
P_{\hat{r}_{-}}\otimes H_{\text{cd}}^{\phi} . \label{OptGate}
\end{align}

\begin{figure}[t!]
	\centering
	\includegraphics[scale=0.45]{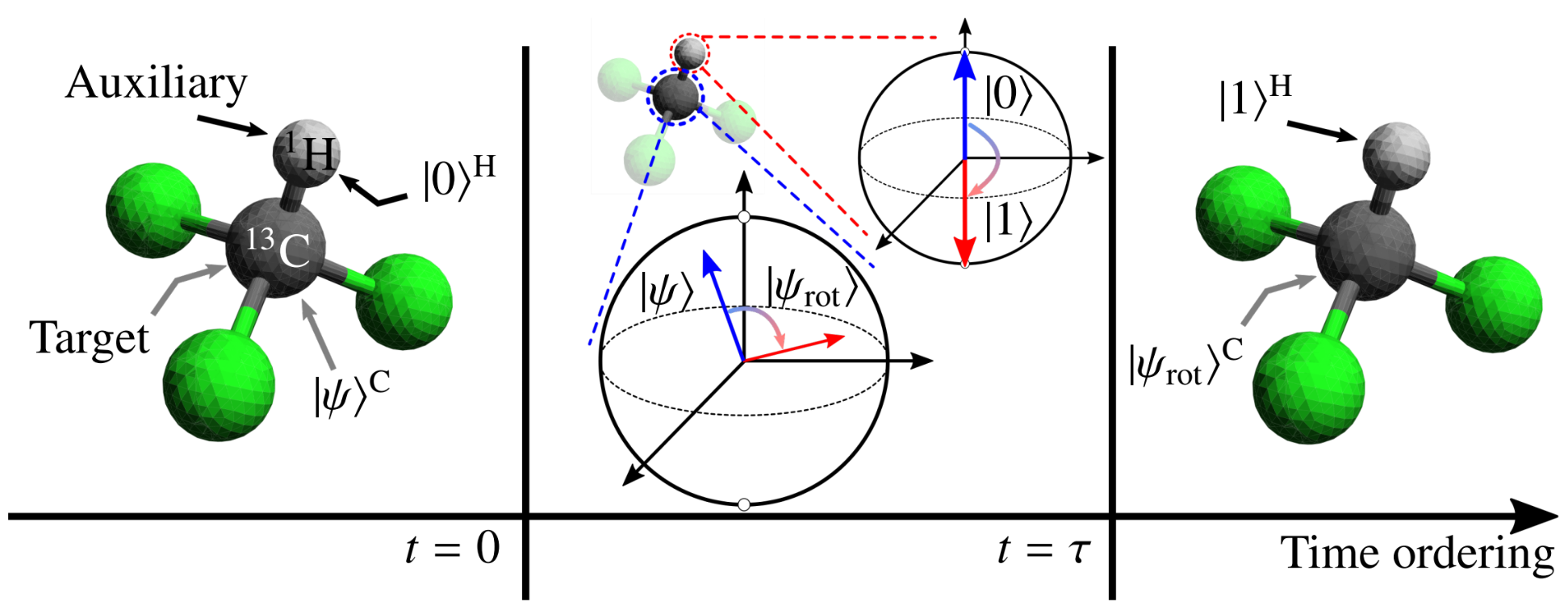}
	\vspace{-0.3cm}
	\caption{Schematic representation of the quantum dynamics for the Chloroform molecule, with the target and ancilla qubits encoded in the Carbon and Hydrogen nuclei, respectively.}
	\label{BlochGate}
\end{figure}

Therefore, the above Hamiltonian can be used to implement any single-qubit rotation. Moreover, one can find time-independent Hamiltonians to implement any $N$-controlled qubit gates~\cite{Santos:18-b}.


The experimental realization of quantum gates is implemented through the Chloroform molecule by taking the $^{13}$C  
nucleus as the target qubit and the $^1$H nucleus as the auxiliary subsystem.  In our experiment, we encode the computational states $\ket{0}$ and $\ket{1}$ into the spin states $\ketus$ and $\ketds$, respectively. The schematic representation of the quantum dynamics 
is illustrated in Fig.~\ref{BlochGate}, where the composite system is initially prepared at $t=0$ and measured in $t=\tau$. Decohering time scales in our system are $T_{1\text{C}}\!=\!7.33$~s and $T_{2\text{C}}\!=\!4.99$~s for $^{13}$C nucleus, $T_{1\text{H}}\!=\!14.52$~s and $T_{2\text{H}}\!=\!0.77$~s for $^1$H nucleus.

We have considered the experimental implementation of the single-qubit phase gate $Z$, which yields the Pauli matrix $\sigma^z$ applied to 
the input qubit, which is prepared in an arbitrary initial state along a direction $\hat{r}$ in the Bloch sphere. 
To this gate, the adiabatic Hamiltonian is
\begin{align}
H_{Z}(s) &= - \omega \left[ \cos (\pi s) \1^{\text{C}} \otimes \sigma_{z}^{\text{H}} + \sin (\pi s) \sigma_{z}^{\text{C}} \otimes \sigma_{x}^{\text{H}}\right] , \label{Hz}
\end{align}
and the standard TQD, and optimal TQD Hamiltonians are given by
\begin{align}
H_{Z,\text{tqd}}^{\text{std}}(s) &= H_{Z}(s) + (\pi/2\tau) \sigma_{z}^{\text{C}} \otimes \sigma_{y}^{\text{H}} , \label{HzTQD}
\\
H_{Z,\text{tqd}}^{\text{opt}} &= (\pi/2\tau) \sigma_{z}^{\text{C}} \otimes \sigma_{y}^{\text{H}} ,  \label{HzOptTQD}
\end{align}
where the subscripts ``C" and ``H" denote operations on Carbon and Hydrogen nuclei, respectively. Eqs.~(\ref{HzTQD}) and (\ref{HzOptTQD}) are obtained from 
Eqs.~(\ref{AdGate}),~(\ref{TradGate}), and~(\ref{OptGate}), respectively, with $\omega\!=\!2 \pi \nu$ and $\nu$ denoting a real frequency. It is worth highlighting that a study on the energy cost to implement adiabatic and TQD Hamiltonians 
has been previously discussed in Ref.~\cite{Santos:18-b} from an operator norm approach. Here, we are interested in analyzing the cost from an alternative point of view, 
where we associate fixed energy amounts for each pulse in a pulse sequence. Therefore, this includes the effective energy spent to implement each pulse of magnetic field, 
while disregarding the free evolution contributions to the quantum dynamics.

\begin{figure}[t!]
	\centering
	\includegraphics[scale=0.57]{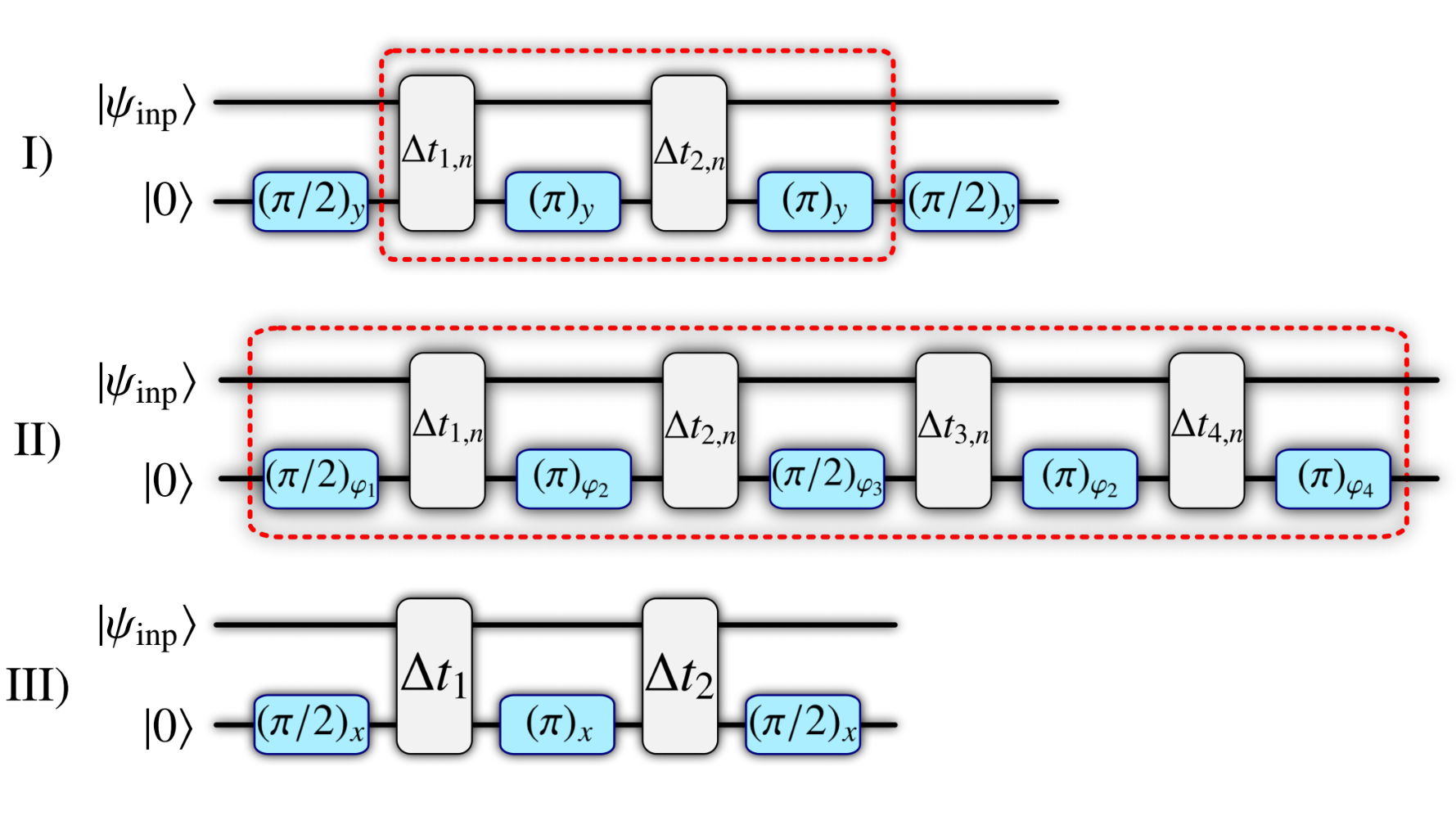}
	\vspace{-0.2cm}
	\caption{Adiabatic, standard TQD, and optimal TQD protocol pulses. Two-qubit blocks labeled with $\Delta t_{j,n}$ and $\Delta t_{n}$ represent free evolution of the chloroform molecule during a time interval $\Delta t_{j,n}$ and $\Delta t_{n}$, respectively, while single-qubit blocks denoted by $(\vartheta)_{\varphi_{n}}$, $(\vartheta)_{y}$ and $(\vartheta)_{x}$ are rotations in Hilbert space of a single qubit. For more details see supplemental material.} \label{Pulses}
\end{figure}

Particularly, in our experimental implementation of the phase gate $Z$, we have set $\nu\!=\!35$~Hz. Differently from the standard TQD and adiabatic Hamiltonians, the optimal TQD protocol provides a time-independent Hamiltonian to realize the phase gate $Z$. In Fig.~\ref{Pulses}, we present the pulse sequences used to implement each Hamiltonian in Eqs.~(\ref{Hz})-(\ref{HzOptTQD}) for any input state. As an illustration, the initial state of the target qubit 
considered in our experiment has been taken as $\ket{\psi(0)} = \ket{+}^{\text{C}}\otimes\ket{0}^{\text{H}}$, with $\ket{+} = (1/\sqrt{2})(\ket{0}+\ket{1})$. 
Each pulse sequence implements the correct dynamics up to a rotation around the $Z$-axis over the auxiliary qubit. 
Since the final state of the auxiliary qubit is $\ket{1}$, the circuits provide the correct output up to a global phase. The dynamics of the system driven by the time-dependent Hamiltonians in Eq.~(\ref{Hz}) and~(\ref{HzTQD}) are implemented by the Dyson series for the corresponding unitaries~\cite{Nielsen:Book}. The pulse composition for the Dyson series is described in details in the Supplemental Material. As shown in Figs.~\ref{Pulses}-I and~\ref{Pulses}-II, the $N$ repetitions are associated to a``trotterization" of the Dyson series for the adiabatic and standard TQD protocols, respectively, with the implementation of 
the Dyson series being exact in the limit $N\rightarrow \infty$. The repeated application of the sequence can lead to the accumilation to experimental sistematic errors, in order to avoid the rrors we have employed NMR composite pulses~\cite{Brown:04}. On the other hand, the optimal TQD Hamiltonian $H_{Z,\text{TQD}}^{\text{opt}}$ can be implemented by using a very short pulse sequence, 
as shown in Fig~\ref{Pulses}-III. 

An immediate discussion arising from the analysis above is the amount of energy required to implement the protocol 
(the pulse sequence associated with the Dyson series). In order to conveniently study the energy cost of implementing the optimal 
TQD protocol, we need to analyze the energy cost of implementing the sequence of pulses able to reproduce the Dyson series for each dynamics.
Consequently, from the pulse sequence presented in Fig~\ref{Pulses}, the energy cost evolving the optimal TQD is (at least) $N$ times less 
than both that associated with the adiabatic and the standard TQD protocols. In fact, if we consider that each pulse in Fig.~\ref{Pulses} [given by a 
rotation $(\vartheta)_{\varphi_{n}}$] is associated with an energy cost $E_{0}$, the overall 
energy cost for implementing the optimal TQD is $E^{\text{opt}}_{\text{tqd}}\!=\!3E_{0}$ (disregarding the energy cost of free evolutions), 
while the energy cost of the adiabatic and the standard TQD are given by $E_{\text{ad}}\!=\!2(N+1)E_{0}$ and $E^{\text{std}}_{\text{tqd}}\!=\!5NE_{0}$, respectively. 
This analysis suggests that generalized TQD can be used to provide Hamiltonians exhibiting optimal pulse sequence.

\begin{figure}[t!]
	\centering
	\includegraphics[scale=0.58]{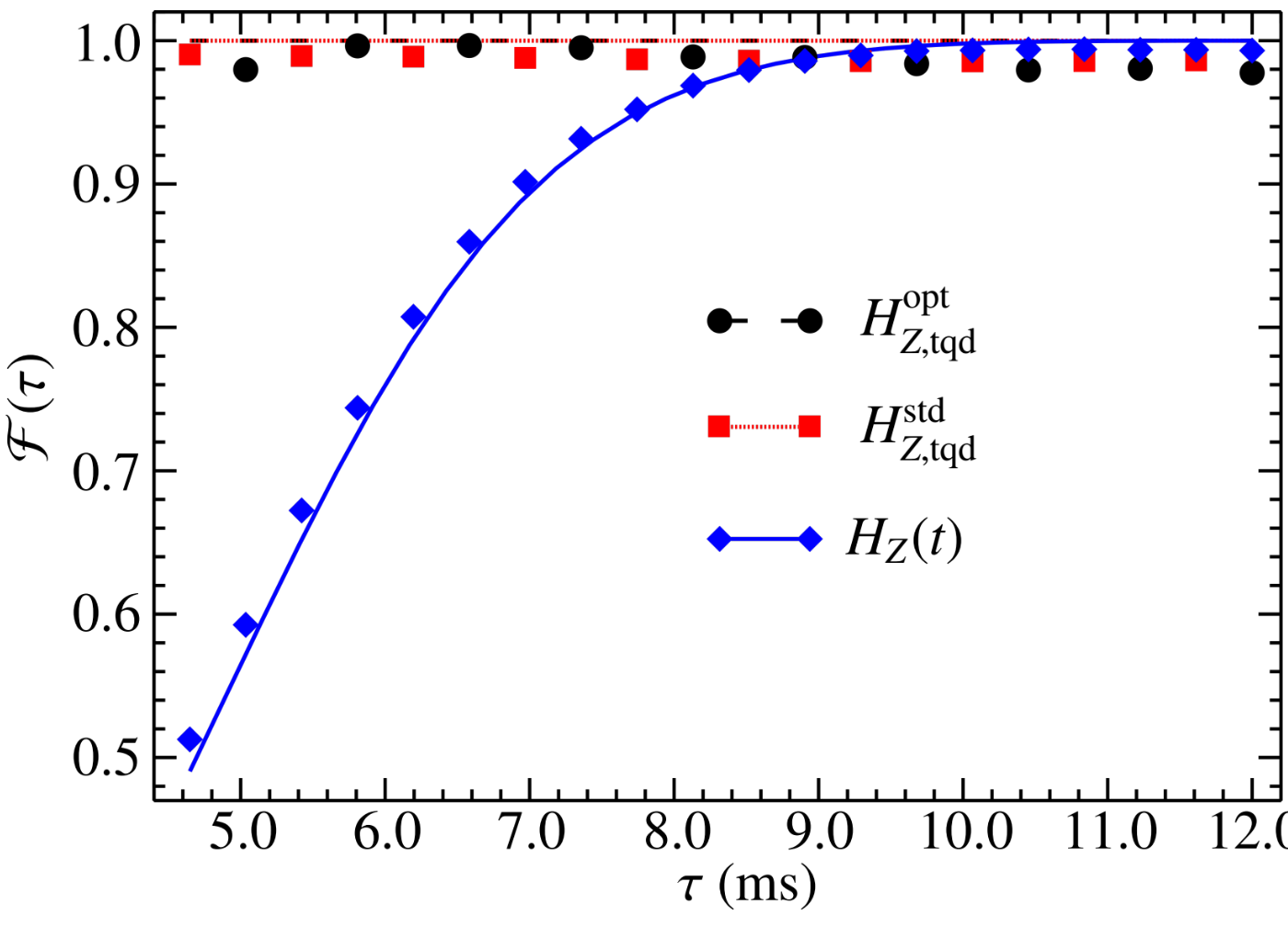}
	\vspace{-0.3cm}
	\caption{Fidelity for the $Z$ phase gate implementation over the initial input state $\ket{+} = (1/\sqrt{2})(\ket{0}+\ket{1})$ encoded in the $^{13}$C nucleus, with the $^{1}$H nucleus representing the auxiliary qubit in the initial state $\ket{0}$. Fidelity $\Fcal(t)$ is computed between the instantaneous ground state of $H_Z(t)$ and the dynamically evolved quantum states driven by the Hamiltonians $H_Z(t)$, $H^{\text{std}}_{\text{Z,tqd}}(t)$, and $H^{\text{opt}}_{\text{Z,tqd}}(t)$. Solid curves and discrete symbols represent theoretical predictions and experimental results, respectively.}	\label{Gate-Fidelity}
\end{figure}


We also consider the fidelity $\Fcal(t)$ between the instantaneous ground state of $H_Z(t)$ and 
the dynamically evolved quantum states driven by the Hamiltonians $H_Z(t)$, $H^{\text{opt}}_{\text{tqd}}(t)$, and $H^{\text{std}}_{\text{tqd}}(t)$. This is 
shown in Fig.~\ref{Gate-Fidelity}. Notice that, as expected for the Hamiltonian $H_Z(t)$, the fidelity improves as the evolution time is increased. 
In contrast, $H^{\text{opt}}_{\text{tqd}}(t)$ and $H^{\text{std}}_{\text{tqd}}(t)$ are capable of achieving high fidelities for any evolution time, since they are 
designed to mimic the adiabatic evolution for arbitrary $\tau$. Notice also that, for the time scale considered in the experiment, which goes up to $\tau = 12$ ms, 
decoherence has little effect, since this $\tau$ is still much smaller than the dephasing relaxation time scale $T_2$.

\section{Conclusion}

In this paper, we have experimentally investigated the generalized approach for TQD in an NMR setup. As a first application, we considered the adiabatic dynamics of 
a single spin$-1/2$ particle in a resonant time-dependent rotating magnetic field. As expected, the adiabatic behavior of the system is drastically affected by the 
resonance phenomenon. On the other hand, the standard TQD and the optimal TQD are both immune to the resonance destructive effect. In particular, 
we have shown that the optimal TQD approach provides a transitionless Hamiltonian that can be implemented with low intensity fields in comparison with the fields 
used to implement the adiabatic and the standard TQD Hamiltonians. As a second application, we have studied adiabatic and counter-diabatic 
implementations of single-qubit quantum gates in NMR. By using a generalized approach for TQD, we have addressed the problem of the feasibility of the 
shortcuts to adiabaticity, as provided by TQD protocols, in the context of quantum computation via controlled evolutions~\cite{Hen:15,Santos:15}. 
By using the generalized TQD Hamiltonian~\cite{Santos:18-b}, we have presented the optimal solution in terms of pulse sequence and resources to 
implement fast quantum gates red through counter-diabatic quantum computation with high fidelity, as shown in Fig.~\ref{Gate-Fidelity}.

The energy-optimal protocol presented here is potentially useful for speeding up digitized adiabatic quantum computing~\cite{Hen:14}.  
Our study explicitly illustrates the performance of generalized TQD in terms of both energy resources and optimal pulse sequence. Since digitized adiabatic quantum computing requires the Trotterization of the adiabatic dynamics, our protocol could be useful in reducing the number of steps used in digital adiabatic quantum processes. 
In addition, such process is independent of the experimental approach used to digitize the adiabatic quantum evolution. 
Generalized TQD may also have impact in speeding up robust holonomic quantum computation~\cite{r1} while keeping energy efficiency. Indeed, 
intrinsic topological properties have been used together with external Hamiltonian decoupling to significantly reduce the influence of noises~\cite{r9} 
(for cancellations of internal noises in the control system, see also Refs.~\cite{r10,r11,r12}.
As a further challenge, the performance of optimal TQD quantum heat engines 
is an unexplored application, both from the theoretical and the experimental point of view. In particular, the proposal of a two-spin system used as a working 
medium in a quantum Otto engine would be an ideal scenario to verify the performance of optimal TQD, since experimentally undesired interactions typically 
arise in the standard TQD implementation of such system~\cite{Baris:18}. These topics are left for future investigation.

\acknowledgments

The authors would like to thank to Prof. Fernando de Melo (CBPF, Brazil) for useful discussions at the initial stage of this work.
A.C.S. is supported by Conselho Nacional de Desenvolvimento Cient\'{\i}fico e Tecnol\'ogico (CNPq-Brazil). A.N. was supported by the Coordenação de Aperfeiçoamento de Pessoal de Nivel Superior - Brasil (CAPES) for the development of this work. M.S.S. is supported by CNPq-Brazil (No. 303070/2016-1) 
and Funda\c{c}\~ao Carlos Chagas Filho de Amparo \`a Pesquisa do Estado do Rio de Janeiro (FAPERJ) (No. 203036/2016). R.S.S., A.M.S., and I.S.O. are supported 
by CNPq-Brazil. AMS is supported by CNPq (304986/2016-0) and  FAPERJ (203.166/2017). The authors also acknowledge
financial support in part by the Coordena\c{c}\~ao de Aperfei\c{c}oamento de Pessoal de N\'{\i}vel Superior - Brasil (CAPES) (Finance Code 001) and by the Brazilian
National Institute for Science and Technology of Quantum Information [CNPq INCT-IQ (465469/2014-0)].

\end{document}